\documentclass[letterpaper]{article} 
\usepackage{aaai25}  
\usepackage{times}  
\usepackage{helvet}  
\usepackage{courier}  
\usepackage[hyphens]{url}  
\usepackage{graphicx} 
\usepackage{natbib}  
\usepackage{caption} 
\nocopyright
\usepackage{algorithm}
\usepackage{algorithmic}
\usepackage{amsmath}

\usepackage{newfloat}
\usepackage{listings}
\DeclareCaptionStyle{ruled}{labelfont=normalfont,labelsep=colon,strut=off} 
\floatstyle{ruled}
\newfloat{listing}{tb}{lst}{}
\floatname{listing}{Listing}
\title{Did You Hear That? Introducing AADG: A Framework for Generating Benchmark Data in Audio Anomaly Detection}

\author {
    Ksheeraja Raghavan\textsuperscript{\rm 1}*, Samiran Gode\textsuperscript{\rm 2}*, Ankit Shah\textsuperscript{\rm 1}*, 
    Surabhi Raghavan\textsuperscript{\rm 3}, Wolfram Burgard\textsuperscript{\rm 2}, Bhiksha Raj\textsuperscript{\rm 1}, Rita Singh\textsuperscript{\rm 1}
}
\affiliations {
    \textsuperscript{\rm 1}Carnegie Mellon University \\
    \textsuperscript{\rm 2}University of Technology Nuremberg \\
    \textsuperscript{\rm 3}University of Pittsburgh \\
}

\usepackage{bibentry}

\begin{document}

\maketitle
\footnotetext[1]{* First three authors have contributed equally to the paper. Correspondence: \{ksheerar, sgode, aps1\}@alumni.cmu.edu}

\begin{abstract}


We introduce a novel, general-purpose audio generation framework specifically designed for anomaly detection and localization. Unlike existing datasets that predominantly focus on industrial and machine-related sounds, our framework focuses a broader range of environments, particularly useful in real-world scenarios where only audio data are available, such as in video-derived or telephonic audio. To generate such data, we propose a new method inspired by the LLM-Modulo \cite{kambhampati2024llm-modulo} framework, which leverages large language models(LLMs) as world models to simulate such real-world scenarios. This tool is modular allowing a plug-and-play approach. It operates by first using LLMs to predict plausible real-world scenarios. An LLM further extracts the constituent sounds, the order and the way in which these should be merged to create coherent wholes. Much like the LLM-Modulo framework, we include rigorous verification of each output stage, ensuring the reliability of the generated data. The data produced using the framework serves as a benchmark for anomaly detection applications, potentially enhancing the performance of models trained on audio data, particularly in handling out-of-distribution cases. Our contributions thus fill a critical void in audio anomaly detection resources and provide a scalable tool for generating diverse, realistic audio data.

\end{abstract}

%

\section{Introduction}
Detecting anomalies is crucial for several reasons like prevention of harm\cite{saligrama2010video}, early detection of unexpected events \cite{cybersec_anomaly} \cite{graph_anomaly}, application in Safety-Critical Systems \cite{anomaly_genera;}, Ensuring data integrity, etc. 
There are many cases where only audio data is available, or is the only usable modality. Thus, it is important to have models trained only on audio data that can detect anomalies. A tragic, yet commonly occurring scenario is that of extortion through phone calls \cite{phone_scam}. In such cases, only audio data are available, and it becomes imperative to find any information that could help solve the crime (e.g. financial crimes). Another common example is that of recordings from phone where the video is not clear, but the audio captures a lot of information. Here again, utilizing only audio data becomes crucial(e.g. Protest videos used by Journalists). Also, with surveillance/CCTV cameras which have the ability to record audio can give information about events not happening in the field of view of the camera. In each of the above cases and many other cases, detecting anomalies and out-of-distribution cases is important. 
In this paper, we take inspiration from the computer vision community to define anomalies \cite{anomaly_survey} \cite{saligrama2010video}, this paper considers only single-scene anomalies.

\begin{figure}[H]
\centering
\includegraphics[height=0.48\textheight]{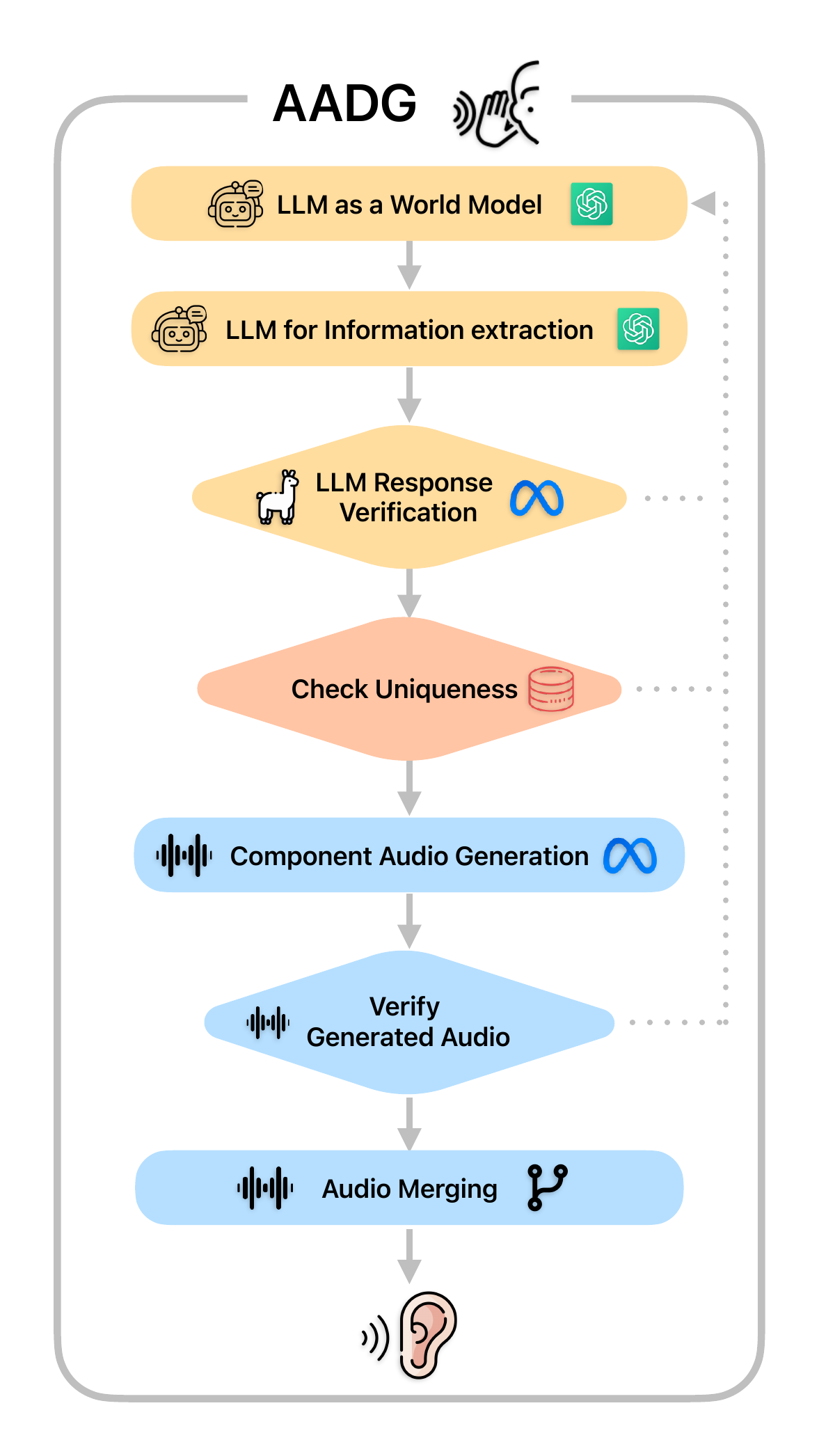} 
\caption{Audio Anomaly Data Generation (AADG), a framework that synthetically generates real life Audio Data with Anomalies by leveraging LLMs as a world model}
\label{Did_you_hear_that}
\end{figure}

\textbf{Definition 1} Audio Anomalies are audio events that stand out within a scene due to their unusual nature relative to the surrounding sounds. This anomaly can arise from the event's position within the audio timeline, its incongruity with the expected auditory context, or the inherent rarity of the sound itself.


There are multiple benchmark datasets for video-based anomaly detection, for instance, Street Scene \cite{ramachandra2020street}, CUHK Avenue\ cite{Cuhk}, ShanghaiTech \cite{shanghaiTech} and UCSD Ped1, Ped2 \cite{ucsdped1}\ cite{ucsdped2}. However, despite the need for audio only anomaly datasets such datasets do not exist. The current use cases of audio only anomaly detection datasets focus solely on industrial and machine data \cite{dcaseDohi2022} \cite{dcaseDohi2022-2} \cite{dcaseHarada2021}. Anomalous data by definition is out of distribution and hence hard to collect since it occurrence is significantly less frequent than general data.

Datasets used to train the current Audio language models and text-to-audio models do not contain complex scenarios and cover a narrow range of scenarios. 
And as a result, audio language models \cite{gama} \cite{agre1987pengi} struggle to perform when faced with complex audio as well as audio that contains anomalies. The current SOTA text to audio models \cite{audiobox} \cite{stable_audio_open}\cite{audiobox} also fails once the prompts start to become descriptive, which is in stark contrast with the current text to image models \cite{stablediffusionpodell2023sdxlimprovinglatentdiffusion}\cite{partiyu2022scaling} and video models \cite{soravideoworldsimulators2024} since they are trained on diverse data.

Collecting audio samples that represent a real-life scene, including an anomaly, is difficult and tedious. Given the importance of audio anomaly detection, data for training and to benchmark is very important, we look towards synthetic generation of such data to augment existing datasets. This is hard since data
that is used to train such models does not contain such scenarios and as a result we must look elsewhere. 

This paper aims to solve this problem by using Large Language Models or LLMs, as a world model \cite{llm_world_model}. Large Language Models such as GPT3.5 \cite{gpt3}, GPT4 \cite{gpt4}, Claude \cite{anthropic2023claude}, LLama \cite{llama} are trained on internet scale data and as a result have text being trained over an unimaginable set of scenarios. Such world models would thus be a great source for creating anomalous and out-of-distribution or less likely scenarios that could happen. We could leverage such language models to help us create audio data. Additionally, current state-of-the-art text-to-audio models \cite{stable_audio_open} \cite{audiobox} fail to generate audio for complex scenarios but are exceptional at creating audios for single cases which are included in their training data and given the correct prompt. This recent advance in both large language models and text-to-audio models could be leveraged to create more complex scenarios.

Recent work by \cite{kambhampati2024llm-modulo} shows that LLMs are indeed great at generating candidate plans but must be verified while planning. And they could be thought to be close to the system 1 \cite{kahneman2011thinking} level of thinking in humans. Our paper aims to utilise this ability of the language models to create scenarios while also verifying their output. We ask the language model to generate a scenario which has anomalies while also being plausible. This output is then used by another call to the LLM to generate the component sounds, the order in which these should be merged. We use a predetermined set of methods of merging these audios and based on the order and the merge suggested by the LLM we create the final audio. We find that the final generated audio outperforms state of the text to audio models for complex prompts and prompts which have anomalies and are out-of-distribution.

Designed with modularity in mind, our framework supports a plug-and-play approach, allowing it to work independently of the language model and the text-to-audio model. The data generated using this framework can thus be used to augment the training data used for text-to-audio models as well as the current audio language models and further improve their audio perception abilities. The data generated using this framework can be used being synthetically generated, providing us access to the exact text description used to generate these data, the component audios, and the timestamps for each audio. This will help both train and benchmark audio anomaly detection models detect the sound events and localize them. This will be the first general-purpose audio anomaly dataset. 

Our key contributions from the work are the following:
\textbf{Novel Framework for Audio Anomaly Data Generation}: The paper introduces AADG (Audio Anomaly Data Generation), a framework that leverages Large Language Models (LLMs) as world models to synthetically generate realistic audio data containing anomalies. This addresses the lack of diverse audio anomaly datasets, especially for real-life scenarios beyond industrial settings. \\
\textbf{Modular and Extensible Approach}: The framework is designed with modularity in mind, allowing for a plug-and-play approach that can work independently of the specific language model and the text-to-audio model used. This flexibility enables the framework to adapt to future advancements in both LLMs and audio generation technologies. \\
\textbf{Creation of the First General-Purpose Audio Anomaly Dataset}: The synthetic data generated by this framework will serve as the first general-purpose audio anomaly dataset, complete with text descriptions, component audios, and timestamps. This dataset will be crucial for training and benchmarking audio anomaly detection models, addressing a significant gap in the field of audio-based anomaly detection.

\section{Related Work}
\textbf{Anomaly detection in video}
Detecting anomalies in videos is quite a common problem \cite{anomaly_survey} \cite{Cuhk} \cite{shanghaiTech} \cite{ucsdped1} \cite{ucsdped2} with multiple datasets for benchmarking. \cite{anomaly_survey} talks about single-scene anomaly detection, and the need to benchmark data for the development of new algorithms. There are multiple such datasets for video. \cite{ucsdped1}, \cite{ucsdped2} are the most widely used\cite{anomaly_survey} and contain videos from different static cameras. \cite{Cuhk} \cite{ramachandra2020street} and \cite{shanghaiTech} are some other common ones, all focusing on single-scene anomalies. Our work aims to introduce similar progress in audio for anomaly detection.
There have been multiple works on the detection of audio anomaly data and can be roughly divided into two parts representation learning and detection methods \cite{singh2023eval}. 


\textbf{Text to Audio generation}
The field of text-to-audio generation has progressed significantly due to innovations in diffusion-based techniques like AUDIT \cite{audit}, AudioGen \cite{audiogen}, and AudioLDM \cite{audioldm} \cite{audioldm2}. Additionally, there have been improvements with auto-regressive models exemplified by AudioGen \cite{audiogen}. AudioGen \cite{audiogen} learns representations from the raw waveform and utilizes a text-conditioned transformer model to produce audio outputs. Moreover, advances in flow matching have further enhanced text-to-audio generation capabilities, as demonstrated in \cite{audiobox}. Recently, \cite{stable_audio_open} introduced a diffusion transformer.

\textbf{LLMs for synthetic data generation} - 
Recent works have shown significant progress in synthetic data generation in the image space in \cite{text_to_image}   \cite{zero_shot_text_to_image} \cite{llama_image_gen} \cite{facegen} etc. However, similar progress has not been seen in the audio space. Zero-shot text-to-image generation approaches have expanded the scope of synthetic data applications by enabling the generation of novel image data from unseen textual prompts, highlighting the model's ability to generalize from limited examples \cite{zero_shot_text_to_image}. Digiface-1m \cite{facegen} dataset exemplifies the practical applications of these technologies, providing a robust framework for testing and improving face recognition algorithms through access to one million digital face images. 
\cite{zero_shot_gen} outlines a method to leverage LLMs to create synthetic datasets produced entirely using pre-trained language models (PLMs) without human interference while emphasizing the efficiency and flexibility of using synthetic datasets to train task-specific models.
\cite{LLM_as)data_generator} explores generation of training data that not only focuses on diversity, but also addresses inherent biases within the data generated by LLMs. It highlights the critical role of using diversely attributed prompts that enhance quality and utility of synthetic datasets improving model performance across NLP tasks. \cite{datadreamer} presents a tool designed to streamline synthetic data generation using LLMs providing a platform to generate, train, and share data sets and models. 

Training on synthetic data can improve the performance of the model \cite{nvidia2024nemotron4340btechnicalreport}. In \cite{dubey2024llama} it has been used to generate training data for text quality classifiers. 

\textbf{LLMs for planning}
Chain-of-thought (CoT) prompting has emerged as a powerful technique to enhance the reasoning capabilities of LLMs by generating intermediate reasoning steps, thereby improving performance on complex tasks such as arithmetic and common sense reasoning. Additionally, the LLM Modulo framework has shown promise in iterative planning and reasoning tasks by establishing a robust interaction between generative models and verifiers, leading to significant improvements in domains like travel planning. These methodologies and frameworks underscore the potential of LLMs to revolutionize automated planning and other complex domains, making them a focal point for future research and development.

\textbf{LLM verification} 
\cite{llm_as_a_judge} investigates the use of GPT-4 as an evaluator of other LLMs, demonstrating the model's ability to assess responses scalably, reducing human involvement, and allowing faster iterations.

\section{Method}
We use a Large Language Model as a world model and prompt it to create a scenario, another call to an LLM further extracts information, such as the summary of the scene, the anomaly, the sound components, the order and how they should be merged. We then use a text-to-audio model to create the component audios and then merge them based on how the LLM instructed. At each stage, we ensure to verify the outputs to be certain the final audio makes sense. In the following, we go through the entire pipeline in detail.

\begin{figure*}[t]
\centering
\includegraphics[width=\textwidth]{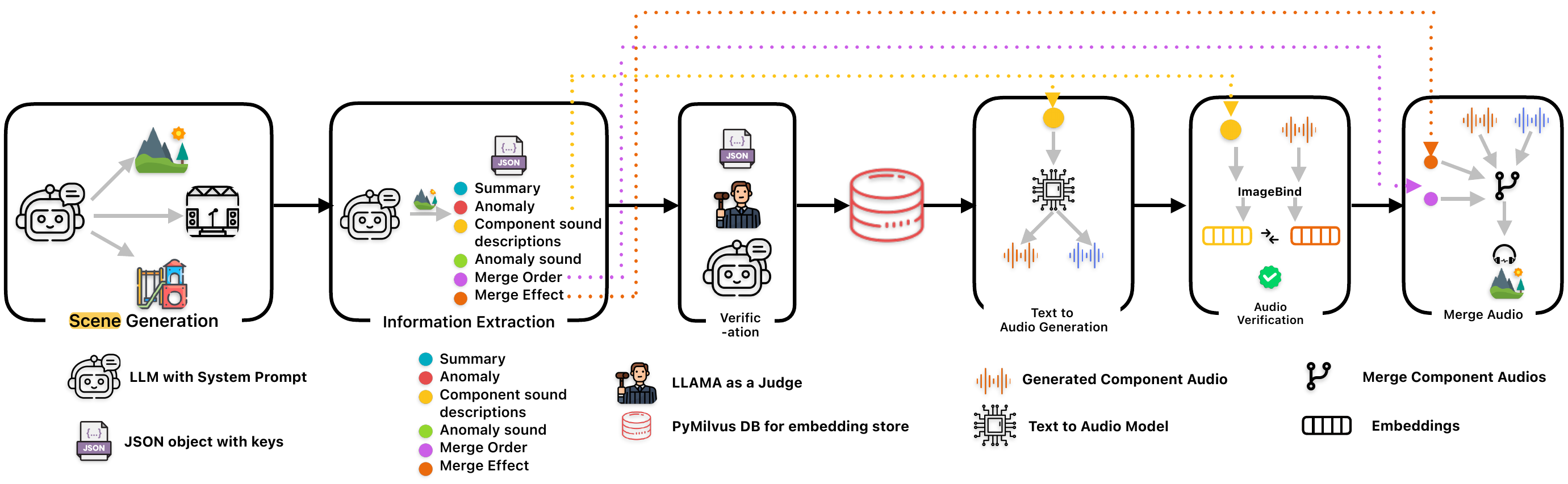} 
\caption{Illustration of the pipeline for generating and verifying anomalous audio data. The process begins with scene generation, followed by information extraction using a Large Language Model (LLM). Individual audio components are synthesized from text descriptions and meticulously verified for accuracy and merged according to LLM instructions, culminating in a dataset of realistic anomalous audio.}
\label{Pipeline}
\end{figure*}

\subsection{Scenario Generation}
Large Language Models have been trained on internet scale data and thus our good world models, they have been known to create great candidate plans \cite{kambhampati2024llm-modulo}. We prompt the LLM to create a scenario which is used to generate scenarios which are plausible in the real world and have sufficient information to create a scene with other component sound and contains an anomaly. We can vary the number of anomalies and the scene by conditioning the prompt. It is also important to note that the way the framework is created, we could also use this same method but change the prompt to create different types of audio. The scenario generally varies according to the size of the language model and the temperature used. We use GPT3.5 and GPT4o for our experiments, but our method is independent of the language model and can, in fact, be replaced by other models as well. We notice that the larger model provides more descriptive and creative outputs and takes particular care about the general scene and what sorts of audio would be generated in such a scene. Additionally, the prompt pushes the model to only consider scenes that would create sounds that would be sufficient enough to distinguish the scene. The number of anomalies can be specified; in practice we have limited it to just one anomaly and find that it is enough to create a benchmark for the current audio language models and is also complicated enough that the current audio language models break. The temperature used to generate such scenarios also affects the final output, we noticed that though an higher temperature was better for the creativity of the model, the GPT4o model broke sometimes with higher temperatures and created non-sensical output with unicode characters. We hypothesize that prompt conditioning the model to generate anomalies and the temperature being high breaks the language model.

\subsection{Information Extraction}
Once we have the initial scenario from the first call to the LLM, we make a second call to the LLM to help us extract useful information and in a format that will be easier for us to use. Through this call, we extract the summary of the scenario, the anomaly present in the scenario. This information will also be useful when these data are used for training or benchmarking evaluation since we have the exact explanation of the audio. Further, this call helps us get the component sounds which we will be generating along with the order in which they would be merged and the type of merge used to merge them. The creative abilities of the LLM allow it to summarize the scenario and extract the information that we need while understanding what we are using this information for and thus providing us what we need for our final audio generation. To force the LLM outputs to be in the format in which we want them, we use the pydantic library \cite{pydantic2024}. This allows us to get the output as dictionaries, which can then be used in the following steps to generate and merge the audios. The component audio information contains the information needed to generate the audio, the order instructs in which order these audios must be used while the merge type tells us how to merge the component audios. This information is crucial and we find that given the scene information the LLM understands what sounds can provide enough information to create the scene and at the same time how these sounds should be merged and in which order should they be merged to create a coherent whole. It also understands what the anomaly is given the scene and also provides information on why that is the anomaly and how and where the anomalous sound should be added to create an audio that sounds realistic.

\subsection{Verification - Language Model}
Large Language Models though creative suffer from multiple issues such as hallucination \cite{hallucination1} \cite{banishing_hallucination} \cite{hallucination_code}, \cite{reasoning2} \cite{reasoning_plan}  lack of reasoning and consistency, redundant outputs \cite{redundancy} \cite{redundancy2} etc. Inspired by LLM Modulo \cite{kambhampati2024llm-modulo} we take advantage of the LLM's impressive creative capabilities but also constrain it by verifying it's outputs to check for issues which might creep in. We check for logical flaws, alignment with the output we want and if the output actually is coherent.

\subsubsection{Logical Verification of Output}
Since the extraction creates outputs that we use in downstream tasks we have to check if they logically make sense. We find that language model fails in number of different ways and thus we try to verify each part. One of the ways it fails is by creating merge types that do not exist in our designated methods, hence we check if the generated merge types lie within what we use. Another way the language model fails and we have to check for is with the number of component sounds, the order and the merge types not being equal. We also find cases where the scenario doesn't make sense and contains nonsensical text. There are also cases where the component audios contain audios that do not make sense with words such as silence, confusion, nervousness etc. These are sounds that need to be checked. 

\subsubsection{LLM as a Judge}

We utilize Llama \cite{llama} as an evaluative tool to verify the responses generated by GPT \cite{gpt4}, by setting up Llama \cite{llama} as an independent impartial judge.  The evaluation uses a Single Answer Grading Framework \cite{llm_as_a_judge} to directly assign a score to the GPT responses.  This setup introduces a layer of quality control, ensuring that the responses generated by GPT align with the prompts that AudioCraft \cite{audiocraft_meta_2024} can use for component audio generation, while eliminating the need for human intervention.

In this study, we leverage Llama  as an impartial evaluation tool to assess the quality of the responses generated by GPT-4 . By employing a Single Answer Grading Framework , Llama directly assigns scores to GPT-4's outputs, introducing a layer of quality control. This approach ensures that the generated responses align with prompts suitable for component audio generation in AudioCraft , while eliminating the need for human intervention. Our methodology showcases the potential of using large language models for automated evaluation, enhancing the efficiency and scalability of the response generation process. The integration of Llama as an independent judge demonstrates a novel approach to quality assurance in natural language generation tasks.

\subsection{Component Audio Generation}
Once we have access to the audio components from the extraction, we pass them to a text-to-audio model that creates the audio components. The advantage of our method is that the text-to-audio models can be replaced as we find better ones, or we could also use multiple text-to-audio models. In practice, we use Audiogen \cite{audiogen}. Audiogen is a textually guided model part of AudioCraft \cite{audiocraft_meta_2024}. We find it to be the best for our case because it is open-source, although there are better models such as Audiobox \cite{audiobox} but they are not available for use. The text-to-audio model can create good audios as long as the prompt isn't complex, in our case, the extraction helps keep the prompts small. However, we could also make the component sound descriptions more informative by conditioning the prompt of the LLM call for extraction. Based on the model that we are currently utilizing, we have determined that utilizing less detailed texts yields better results in practice.

\subsection{Verification - Audio Generation}

The text-to-audio model, much like large language models (LLMs), isn't perfect and often fails to faithfully align with the text prompt used to generate the audio. The model is trained on a dataset that might not comprehensively cover the full spectrum of possible sounds, leading to out-of-distribution (OOD) cases. For instance, we observed that the model performs better on prompts such as "cat meowing" compared to "lion roaring," likely because we have seen more examples of the former. Additionally, prompts involving timing specifications, such as "periodic announcement prompts," frequently confuse the model. The model also struggles with accurately rendering conversations, unless they are in the background, where it tends to work better.

To ensure that the final audio semantically aligns with the text, we propose using a multimodal model trained to align embeddings representing the same scenario across different modalities. We utilize ImageBind \cite{girdhar2023imagebind}, though other models such as Audio CLIP could also be considered. Since ImageBind has been trained contrastively to align the embeddings of the same label across different modalities, the output embeddings of ImageBind for the generated audio and the text prompt should be close according to a chosen distance metric. In our case, we used cosine similarity to measure the alignment between the text and audio embeddings. Formally, given the text embedding $\mathbf{E}_{\text{text}}$ and the audio embedding $\mathbf{E}_{\text{audio}}$, we compute the cosine similarity as follows:

\begin{equation}
\text{Cosine Similarity} = \frac{\mathbf{E}_{\text{text}} \cdot \mathbf{E}_{\text{audio}}}{\|\mathbf{E}_{\text{text}}\| \|\mathbf{E}_{\text{audio}}\|}
\end{equation}

For each generated audio, if the cosine similarity is above a predefined threshold, we accept the audio as semantically aligned. However, in practice, we found that ImageBind did not perform well with the generated audio even when the audio sounded accurate. Despite this, the similarity score was significantly lower (by an order of magnitude) for semantically dissimilar audios compared to semantically similar ones. To enhance the verification process, we apply a sigmoid regularizer to make the differences more pronounced:

\begin{equation}
    \text{Regularized Similarity} = \sigma(\alpha \cdot \text{Cosine Similarity} - \beta)
\end{equation}

Where $\sigma(x) = \frac{1}{1 + e^{-x}}$ is the sigmoid function and $\alpha$ and $\beta$ are tunable parameters that control the scaling and shift. We then only accept audios whose regularized similarity is above a certain threshold, improving the robustness of the verification.

\subsection{Audio Merging}
Once we have verified the component audios, we then proceed to merge them. The order and the way of merging these audios depend on the instructions provided by the language model. There are several possible merge types to choose from, including cross-fade, overlay, fade-in, and fade-out. We carry out the merging process according to the order specified by the language model, with each audio merging with the next based on the suggested method. During this process, we ensure that the audios are normalized to maintain consistent audio levels. These merge types can be expanded, although we have found these to be sufficient in practice. As we merge the audios in order, each new audio is joined at the end of the previous one. For example, if it is a fade-in, the new audio fades in from the end of the previous merged audio. If it's a fade-out, the new audio is added to the end of the previous merged audio, and the end of the new audio fades out. Similarly, in a cross-fade, the previous merged audio cross-fades into the new audio. If it's an overlay, the alignment of the new audio depends on its length relative to the previous merged audio. Simultaneously, we store the timestamps of each audio in the final audio, which can be utilized for evaluation, training models for anomaly detection, or event detection. It is important to note that the merging methods can be altered by adding new methods or utilizing a new learned model, such as using flow matching.

\subsection{Final Data}
The final data thus contains the audio components, the merged audio itself and metadata which contains the scenario, the summary of the scenario, the anomaly, a description of why it is anomalous, the text description of the audio components, the order in which the audios have been merged, the method with which they have been merged and the time stamps of the component audios in the final audio.


\section{Evaluation}
To demonstrate the usefulness of our model to the research community, we illustrate how existing models utilizing audio could be enhanced by accounting for anomalous scenarios and using synthetic data during training.


\subsection{Comparison against State of the Art Text to Audio generation models}
We show that the data generated from our framework is better than those generated from the state-of-the-art text to audio generation models, especially when the prompt used to generate them is complex or contains anomalies. We compare against Audiobox\cite{audiobox} and Stable Open Audio \cite{stable_audio_open}.

The vocabulary set used by models like Audiobox and Stable Open Audio is inherently limited because they are trained on datasets like AudioSet, which only cover a specific range of sounds and scenarios. This means that these models can generate audio based on a fixed set of categories and sound events that are well represented in their training data, such as common environmental noises, simple musical notes, or speech sounds. However, when the prompt involves complex or uncommon audio scenarios, these models fail to generate accurate outputs because they do not have the vocabulary or understanding required to handle sounds outside of their limited training set. This restriction in their sound vocabulary leads to repetitive or irrelevant audio output when faced with prompts that require a broader or more specialized range of sounds, limiting their effectiveness for tasks like anomaly detection, where nuanced and varied sounds are crucial. This gap underscores the need for a more flexible framework that can simulate a wider variety of audio events beyond the limited vocabulary of existing models.

\begin{table}[t]
\centering
\begin{tabular}{l|l}
\hline
    \textbf{Model} & \textbf{PREF} \\ \hline
    Stable Audio Open & 0.12 \\
    \textbf{Ours AADG} & \textbf{0.88} \\
    \hline
\end{tabular}
\caption{Comparing the adherence to the text prompt by text to audio models when the prompts are complex and introduce anomalies}
\label{text-to-audio-table}
\end{table}

\subsection{Benchmarking Audio Language Models}

Current Audio Language Models (ALMs) are primarily trained on datasets devoid of anomalies, such as clean speech with minimal background noise. This raises concerns about their robustness and ability to handle real-world audio complexities.  To assess the true capabilities of state-of-the-art ALMs, we tested using the current state-of-the-art \cite{gama} asking it to predict complex audios. We randomly sample audio and ask participants to choose the how accurate the description is for the audio. We ask them to score the model from 1 to 5, 5 being extremely accurate and 1 being inaccurate. We report the Mean Opinion Score(MOS) of the participants.
We find that the Audio Language Model can understand the easier audios but fails to completely comprehend the complex audios. This is likely because the model was not trained on such data. Models like GAMA\cite{gama} are impressive, but will become much better once trained on datasets augmented with complex data. 

\begin{table}[t]
\centering
\begin{tabular}{l|l}
\hline
    \textbf{Model} & \textbf{MOS} \\ \hline
    GAMA against simple audios & 4.00 \\
    GAMA against complex audios & 3.21 \\
    \hline
\end{tabular}
\caption{Evaluating the audio understanding abilities of current state-of-the-art audio language model. Mean opinion score higher the better}
\label{text-to-audio-table}
\end{table}

\subsection{Comparing Against Audio Separation Models}
Audio separation models \cite{audiosep_interspeech} \cite{audiosep_separate} separate a component audio from a test audio that has more than one sound present. It does so with the supervisory help of a prompt explaining what the audio is. Audio separator models much like the other models we have talked about don't perform well when trying to separate audios from complex or anomalous audios. We believe this is because when faced with audios containing parts which it hasn't seen during training or when dealing with audios with multiple smaller audios merged together, it doesn't completely understand which component audio is separate from the others. We wanted to show through an objective metric that complex audios and their corresponding prompts break such models. As part of our dataset, we have access to the component audio description along with the prompt and the component audios, among other things. Audio separator models like those introduced in \cite{audiosep_interspeech}\cite{audiosep_separate} should, ideally, be able to separate our model into it's components. And, hence in such cases, the separated audios extracted using our text description should align with the original audio generated using the same prompt. However, we suggest that if the audio has anomalies and is complex or has audios that are out of distribution the separation will perform poorly compared to when the audios are simple. To test this, we generate audio samples that are generated using simple and complex prompts, and we find that the ones generated with complex prompts are harder to separate\ref{fad_table}. We use Frechet Audio Distance(fad) similar to \cite{audiobox}\cite{audiogen} to find if the separated and orignal audios are close and we find that for complex audios the FAD increases.
\begin{table}[t]
\centering
\begin{tabular}{l|l}
\hline
    \textbf{Prompt Complexity} & \textbf{FAD} \\ \hline
    Lower Complexity & 5.5015 \\
    Higher Complexity & \textbf{6.775} \\
    \hline
\end{tabular}
\caption{Comparing the audio separation capabilities for different levels of prompt complexity. The lower the FAD score, the better the match}
\label{fad_table}
\end{table}

\section{Limitations}

Anomaly detection in audio presents a major challenge due to the limited vocabulary set used by current state-of-the-art text-to-audio models such as Audiobox and Stable Open Audio. These models are trained on datasets like AudioSet, which only encompass a specific range of sounds and scenarios. Consequently, they struggle to produce accurate audio when presented with complex or uncommon audio scenarios that fall outside their training data. This restricted sound vocabulary results in repetitive or irrelevant output for anomalous audio prompts. Our approach, AADG, aims to address this limitation by generating out-of-vocabulary sounds with complex real-life descriptions. However, the generated audio may sometimes sound unnatural, which can be computationally intensive to detect. The inherent limitation of the current audio generation models starts to degrade in audio quality for very long duration audio generation, which limits the audio duration the AADG framework will be able to generate for complex anomaly description in a given acoustic scene. 


\section{Conclusion}

The proposed framework for generating benchmark data in audio anomaly detection addresses significant gaps in the field by providing a more versatile and scalable approach compared to existing methods. Unlike traditional datasets that focus narrowly on industrial or machine-related sounds, this framework uses large language models (LLMs) to simulate a broader range of real-world scenarios, making it particularly useful in contexts where only audio data are available, such as surveillance or telephonic recordings. The modular design allows for flexibility in integrating different LLMs and text-to-audio models, enabling the generation of complex scenarios with anomalies that are difficult to capture in real-world data. While current text-to-audio models still face challenges with generating realistic audio for complex prompts and anomalies, the framework introduces multi-stage verification processes to minimize logical flaws, misalignment, and inconsistent outputs. Additionally, by using multimodal models like ImageBind for verification, the framework improves the reliability of the generated data, although this process still has limitations in handling certain out-of-distribution cases. Overall, our approach fills a critical gap in the field by providing a scalable tool for creating diverse and realistic audio datasets, which are essential to advance audio anomaly detection technologies.

\section{Acknowledgements}
This work was partly supported by Top Professorship as part of the High Tech Agenda (Spitzenprofessur)

\bibliography{aaai25}

\end{document}